\documentclass[journal=jpclcd,manuscript=letter,layout=onecolumn,longbibliography]{achemso}
\usepackage[T1]{fontenc}
\usepackage{geometry}
\usepackage{color}
\usepackage{float}
\usepackage{units}
\usepackage{mathtools}
\usepackage{amsmath}
\usepackage{amsthm}
\usepackage{amssymb}
\usepackage{graphicx}
\usepackage{setspace}
\usepackage{ulem}
\makeatletter

\@ifundefined{date}{}{\date{}}

\usepackage{hyperref}
\hypersetup{
    colorlinks,
    citecolor=blue,
    filecolor=blue,
    linkcolor=blue,
    urlcolor=blue
}

\usepackage{notoccite}

\usepackage {multicol}

\usepackage{xcolor}
\definecolor{forestgreen}{RGB}{34,139,34}
\definecolor{orangered}{RGB}{239,134,64}
\definecolor{darkblue}{rgb}{0.0,0.0,0.6}

\usepackage[final]{pdfpages}

\makeatother

\author{H. da Silva Jr.}
\affiliation{Department of Chemistry and Biochemistry, University of Nevada, Las Vegas, Nevada 89154, USA}

\author{B. K. Kendrick}
\affiliation{Theoretical Division (T-1, MS B221), Los Alamos National Laboratory,
Los Alamos, NM 87545, USA}

\author{H. Li}
\affiliation{Department of Physics, Temple University, Philadelphia, Pennsylvania
19122, USA}

\author{S. Kotochigova}
\affiliation{Department of Physics, Temple University, Philadelphia, Pennsylvania
19122, USA}

\author{N. Balakrishnan}
\email{naduvala@unlv.nevada.edu}
\affiliation{Department of Chemistry and Biochemistry, University of Nevada, Las Vegas, Nevada 89154, USA}

\title[An \textsf{achemso} demo]
{Non-adiabatically driven quantum interference effects in the ultracold K + KRb $\longrightarrow$ Rb + K$_{2}$ chemical reaction}

\begin{document}

\begin{tocentry}
\includegraphics[width=5cm, height=5cm]{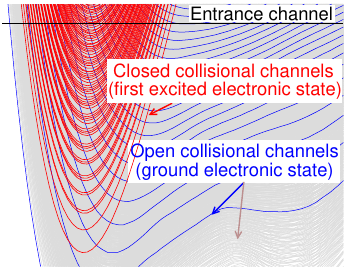}
\end{tocentry}

%
%

\begin{abstract}
The K + KRb $\longrightarrow$ Rb + K$_{2}$ chemical reaction is
the first ultracold atom-diatom chemical reaction for which experimental
results have been reported for temperatures below 1 $\mu$K more than
a decade ago. The reaction occurs through coupling with an excited
electronic state that is accessible even in the ultracold limit. A
previous quantum dynamics study, excluding non-adiabatic effects,
has reported a rate coefficient that is about 35\% below the experimental
value. Here, we report the first non-adiabatic quantum dynamics study
of this reaction and obtain rate coefficients in better agreement
with experiments. Our results show that short-range dynamics mediated
by coupling with the excited electronic state introduces quantum interference
effects that influence both the state-to-state rate coefficients and
the overall reaction rates.
\end{abstract}

Cold and ultracold molecules offer a new paradigm for the investigation
of chemical reactions in a regime where quantum effects dominate and
quantum interference effects strongly influence state-to-state chemistry~\cite{krems2008,Carr2009,kendrick2015,balakrishnan2016,Bohn2017,hu2019,Liu2021,kendrick2021,Liu2022}.
In the ultracold $s$-wave regime, where only the $\ell=0$ angular momentum
partial wave contributes, the interference effect can be purely constructive
or destructive, maximizing or minimizing the reaction outcome. It
is also in this regime where subtle effects such as the geometric
phase (GP) plays a crucial role, as
demonstrated in several benchmark studies of ultracold chemistry involving
H + H$_{2}$ and its isotopic counterparts\cite{Kendrick_2015_PRL,hazra2016,kendrick2016a,kendrick2016b,kendrick2018b,kendrick2019}
as well as the O + OH ~\cite{kendrick2015,Hazra_JPCA_2015}  and, most recently, H + O$_{2}$~\cite{wang2025} chemical reactions. In these cases, the GP introduces a sign change to the interference
term among components of the wavefunctions that do encircle the conical intersection (CI)
between two electronic states and those that do not, making constructive
interference destructive and vice versa. The K + KRb$\left(X^{1}\Sigma^{+},\upsilon=0,j=0\right)\longrightarrow$ K$_{2}\left(X^{1}\Sigma_{g}^{+}\right)$ + Rb
chemical reaction is an ideal case of non-adiabatic dynamics in the
ultracold regime where the first excited electronic state of the complex, $a^{2}B^{\prime}$, is directly accessible
in collisions of rovibrationally ground state KRb molecules. Thus, a
dynamical calculation on the lowest adiabatic electronic state  within the Born-Oppenheimer (BO) approximation, with corrections accounting for the
GP effects, cannot fully capture the effect of coupling to the excited
state for this reaction. Rather, it requires a fully non-adiabatic treatment as recently demonstrated for the Li + LiNa $\to$ Li$_2$ + Na reaction~\cite{kendrick2021}.

The first and only measurement of the rate coefficients for the title reaction
was reported at a temperature below 1 $\mu$K by Ospelkaus \textit{et
al.} in 2010 \cite{ospelkaus2010a}. Both the K atoms and KRb molecules
were prepared in their lowest hyperfine states, \textit{i.e.}, $\left|F=\nicefrac{9}{2},m_{F}=-\nicefrac{9}{2}\right\rangle $ for the K atom 
and $\left|m_{\rm{K}}=-4,m_{\rm{Rb}}=\nicefrac{3}{2}\right\rangle$ for KRb, respectively.
As a consequence, the reactive pathway, whose exoergicity
is of about 217 cm$^{-1}$, is the only decay mechanism energetically accessible. Ospelkaus \textit{et al.} measured the
decay rate of the KRb molecules as a function of the K atom density
and extracted a rate coefficient of $1.7\pm0.3\times10^{-10}$ cm$^{3}$/s
without specifying the actual temperature value. However, only
\textit{s}-wave collisions are expected below 1 $\mu$K, obeying the
Wigner threshold regime where the rate coefficient is expected to be independent
of the temperature. In the same work they have also reported for the
first time the rate coefficient for the KRb + KRb $\longrightarrow$
K$_{2}$ + Rb$_{2}$ reaction through a similar analysis but in this
case the measured value corresponds to a specific kinetic temperature
of 250 nK. More detailed experiments of the KRb + KRb reaction with
quantum-state resolution of the K$_{2}$ and Rb$_{2}$ products were
recently reported by Ni and co-workers\cite{hu2019,Liu2021}. Further,
Liu \textit{et al.} have recently  demonstrated that quantum coherence is preserved
when a pair of ultracold KRb molecules, entangled in their nuclear spin states, undergoes
chemical reaction leading to the K$_{2}$ + Rb$_{2}$ products~\cite{Liu_KRb_2024}.
Unfortunately, the KRb + KRb and similar dimer-dimer systems are beyond the scope of current
computational capabilities using a full quantum mechanical formalism.

For the K + KRb chemical reaction the only prediction based on explicit calculations is that of
Croft \textit{et al.}\cite{croft2017a} who reported full-dimensional
quantum scattering calculations on the ground electronic state $X^{2}A^{\prime}$ of
the K$_{2}$Rb system for the total angular momentum quantum number
$J=0$. The predicted rate coefficient at 250 nK, $1.1\times10^{-10}$
cm$^{3}$/s, is about 35\% below the measured value of Ospelkaus and
co-workers. The computations did not account for the GP effects induced by the first excited electronic
state, whose correlated scattering channels are energetically accessible
in collisions with the ground-state reactants. In what follows, we
refer to these calculations as BO-NGP or simply NGP, for no-geometric phase effects.
The specific goal of the present study is to examine to what extent the first excited electronic
state plays a role in the description of the K + KRb$\left(X^{1}\Sigma^{+},\upsilon=0,j=0\right)\longrightarrow$
K$_{2}\left(X^{1}\Sigma_{g}^{+}\right)$ + Rb chemical reaction in
the ultracold regime, and whether its inclusion in the calculations can improve agreement with the experiment. The overall goal is to gain insights into chemistry mediated through coupling to an excited electronic state in a regime where  quantum effects dominate and only a single angular momentum partial-wave contributes to the measured rate coefficient. While substantially more computationally
demanding, we do obtain results in closer agreement with the measured
value. As such, the computations reported here demonstrate
that non-adiabatic corrections may play an important role in ultracold
chemical reactions and that quantum interference effects mediated
by the non-adiabatic dynamics can control chemistry at ultracold temperatures.

Non-adiabatic dynamics has largely been modeled using time-dependent wave packet (WP) methods.~\cite{yarkony1996,domcke2012,tully2012,guo2016}
 The WP methods are, however, somewhat inefficient, and to some extent less accurate, in the ultra-low range of kinetic energies\cite{huang2018}.
 Here we adopt a novel non-adiabatic quantum reactive scattering formalism developed by Kendrick for the problem of two electronic states, and
 based on a time-independent coupled-channel (CC) formalism in hyperspherical coordinates\cite{kendrick2018a,kendrick2018b,kendrick2019,kendrick2021}.
The 2$\times$2 diabatic representation of the Schr\"odinger equation at a given collision energy, $E_{\mathrm{coll}}$, is written as

\begin{equation}
  \label{eq:DiabaticSchrodingerEquation}
  \begin{array}{c}
    \left[
      \left(
        \begin{array}{cc}
            \mathbf{T}_{1} & \mathbf{0}
            \\
            \mathbf{0}     & \mathbf{T}_{2}
        \end{array}
      \right)

      +

      \left(
        \begin{array}{cc}
          \mathbf{V}_{11} & \mathbf{V}_{12}
          \\
          \mathbf{V}_{21} & \mathbf{V}_{22}
        \end{array}
      \right)

      - E_{\mathrm{coll}}\mathbf{I}
    \right]

    \left(
      \begin{array}{cc}
        \psi_{1} \\ \psi_{2}
      \end{array}
    \right)

    = \mathbf{0}
  \end{array},
\end{equation}

\noindent where $\mathbf{I}$ is the identity matrix, $\mathbf{T}_{n}=-\hbar^{2}\nabla_{n}/2\mu$
is the matrix representation of the kinetic operator with $\nabla$ denoting the derivative with respect to the radial and angular coordinates,
$\mu$ is the three-body reduced mass, and $\psi_{n}$ is the scattering solution matrix for the $n$-th electronic state, $n=1,2$ in the present
case. The diabatic potential is the block off-diagonal matrix, whose elements are transformed from usual adiabatic potential energy surfaces (PESs),
denoted here $V_{1}$ and $V_{2}$.

Unlike the Born-Oppenheimer approximation, in which electronic states remain fairly separated in energy during a collision, \textit{i.e.},
the electronic and nuclear coordinates are adiabatically decoupled, Eq. (\ref{eq:DiabaticSchrodingerEquation}) is valid
whenever the nuclear motion can induce an electronic (de)excitation, but without the need for double-valued boundary conditions\cite{higgins1958,herzberg1963,obrien1964,kendrick2018a}.
In particular, its block-diagonal kinetic energy term is free of non-adiabatic couplings,
which are notoriously difficult to evaluate numerically. As discussed at length by Kendrick\cite{kendrick2018a,kendrick2018b,kendrick2019,kendrick2021},
Eq. (\ref{eq:DiabaticSchrodingerEquation}) can be rewritten as a set of CC equations, yielding a numerical framework similar to that of a NGP calculation.

The method is implemented in the APH3D suite of programs, which has also found extensive application in describing a variety of systems
through conventional Born-Oppenheimer calculations \cite{kendrick2000,kendrick2001,kendrick2003a,kendrick2003b,makrides2015,kendrick2016a,kendrick2016b,hazra2016,croft2017a,croft2017b,silva2022a,silva2023,morita2023}.
An in-depth discussion of the hyperspherical CC equations and numerical strategies to solve them, as implemented on APH3D, has been given by
Kendrick and co-workers on many occasions and is omitted here for brevity. For the sake of the discussion below,  a brief comment will suffice on key aspects of the calculations.
The hyperradius, $\rho$, describing the simultaneous radial atom-diatom
relative motion in all arrangements is partitioned into an inner region,
using the adiabatically adjusting principal axis hyperspherical (APH) coordinates of Pack and Parker\cite{pack1987},
where collision-induced rearrangement is more likely to occur. In
the outer region, where the different atom-diatom arrangement channels
are largely decoupled, Delves hyperspherical coordinates\cite{delves1958,delves1960}
are employed. The set of radial coupled equations is solved using Johnson\textquoteright s
log-derivative method \cite{johnson1973}, first from $\rho_{\mathrm{min}}$
to $\rho_{\mathrm{match}}$. At $\rho_{\mathrm{match}}$ the numerical
solutions from the outermost sector of the APH region are projected
onto solutions at the innermost sector of the Delves region. The propagation
is then continued from $\rho_{\mathrm{match}}$ to $\rho_{\mathrm{max}}$, where scattering
boundary conditions are applied, yielding a scattering matrix\cite{pack1987}.

The PES for the ground electronic state $X^{2}A^{\prime}$ of the K$_{2}$Rb system was first reported in the original work of Croft \textit{et al.}~\cite{croft2017a}. To construct the diabatic PESs we use an approach similar to that we employed for the  Li + LiNa ultracold
chemical reaction~\cite{kendrick2021}. The diabatic PESs are obtained from the adiabatic pair corresponding to the lowest two electronic states, whereas the non-adiabatic
coupling term (NACT) between them is evaluated using the strategy described by Baer.~\cite{Baer2002}.

In these new calculations,
the K$_{2}$Rb complex was modeled as an effective system of three valence
electrons moving in the field of atomic ions represented by effective core potentials (ECP), including relativistic scalar effects. Namely, the ECP18SDF is used for
the K$^{+}$ core and ECP36SDF for Rb$^{+}$. In addition, the ECPs are complemented with core polarization potentials (CPPs), to account for electronic correlations between core
and valence electrons.~\cite{Fuentealba1983_1,Fuentealba1983_2} The basis sets used to describe the valence electrons associated with the ECPs were in their uncontracted form. Finally,
$spdf$ diffuse functions with those exponents reported by \.{Z}uchowski and co-workers~\cite{Zuchowski2010} were added.

Multi-configurational self-consistent field (MCSCF) calculations were performed to obtain the reference determinants for a subsequent multi-reference configuration
interaction (MRCI) step, resulting in the adiabatic PESs for the lowest two electronic states of K$_{2}$Rb, \textit{i.e.}, $V_{1}\left(\bf{x}\right)$ and $V_{2}\left(\bf{x}\right)$, where $\bf{x}=\left(r,R,\theta\right)$ is a set of Jacobi coordinates. The calculations have been performed
on a 13-point Gauss-Legendre quadrature grid for $\theta$, between $0^{\circ}$ and $180^{\circ}$, and radial grids from 3.2 \textit{a.u.} to 10 \textit{a.u.},
with a constant spacing of 0.2 \textit{a.u.}, and from 10 \textit{a.u.} to 19 \textit{a.u.} at steps of 0.3 \textit{a.u.}, for $R$ and $r$ respectively. This leads to
more than $50\, 000$ discrete spatial data points.

All \textit{ab initio} calculations were performed with the Molpro software package~\cite{Werner2012}. At every spatial grid point, the NATC matrix was evaluated with a
numerical finite difference methodology (the DDR procedure as implemented in Molpro~\cite{Werner2012}) and then numerically integrated to obtain the adiabatic-to-diabatic transformation
(ADT) angles, $\beta$. Thereby, we construct the diabatic potentials: $V_{11} = V_{1}\cos^{2}\left(\beta\right) + V_{2}\sin^{2}\left(\beta\right)$; $V_{22} = V_{2}\cos^{2}\left(\beta\right) + V_{1}\sin^{2}\left(\beta\right)$;
and $V_{12} = V_{21} = V_{2}\cos\left(\beta\right)\sin\left(\beta\right) - V_{1}\cos\left(\beta\right)\sin\left(\beta\right)$.~\cite{kendrick2018a}

\begin{figure}[H]
\begin{centering}
\includegraphics[scale=1.2]{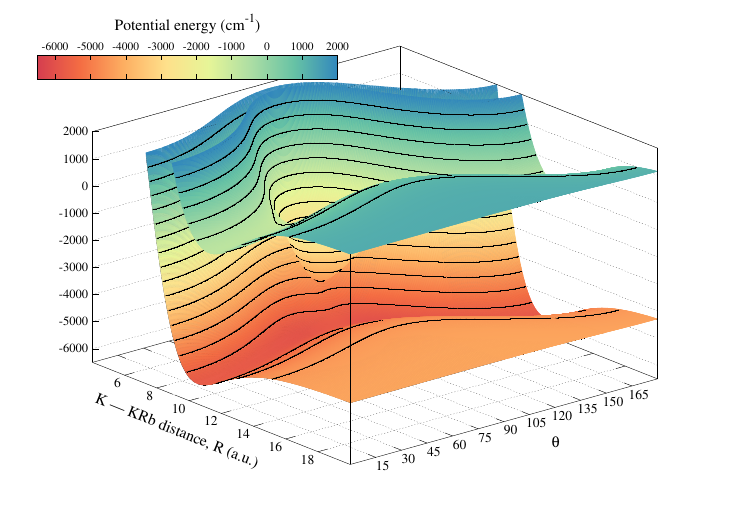}
\par\end{centering}
\caption{\label{fig:AdiabaticPES} Global PESs V$_{1}$ (bottom) and V$_{2}$ (top), in cm$^{-1}$, for the K + KRb$\left(X^{1}\Sigma^{+}\right)$ arrangement as functions of the Jacobi coordinates $\left(R, \theta\right)$
at a fixed KRb internuclear distance of $r$ = 7.7 atomic units. The same energy scale and color coding are used on both surfaces. Isolines varying every 500 cm$^{-1}$ from -6500 to 2000 cm$^{-1}$ are displayed.}
\end{figure}

Finally, the global diabatic PESs and the off-diagonal coupling surface are constructed by interpolating the discrete electronic structure data points using the reproducing kernel
Hilbert space (RKHS) approach.~\cite{Hollebeek1997, Unke2017}. The short-range part of both adiabatic PESs for the K + KRb$\left(X^{1}\Sigma^{+}\right)$ arrangement, V$_{1}$ and V$_{2}$, are presented in Fig. (\ref{fig:AdiabaticPES}) as
functions of the Jacobi coordinates $\left(R, \theta\right)$ at a fixed KRb internuclear distance of $r$ = 7.7 \textit{a.u.} The same energy scale and color coding are used to represent isoenergetic regions of both surfaces. Along the collinear geometries, the incoming K atom approaches on the Rb side of the molecule at $\theta = 0^{\circ}$,
and on the second K atom at $\theta = 180^{\circ}$. An inspection of Fig. (\ref{fig:AdiabaticPES}) shows how the upper PES corresponding to the first excited electronic state deepens toward the lower PES
in the vicinity of C$_{2v}$ geometries.

\begin{figure}[H]
\begin{centering}
\includegraphics[scale=0.4]{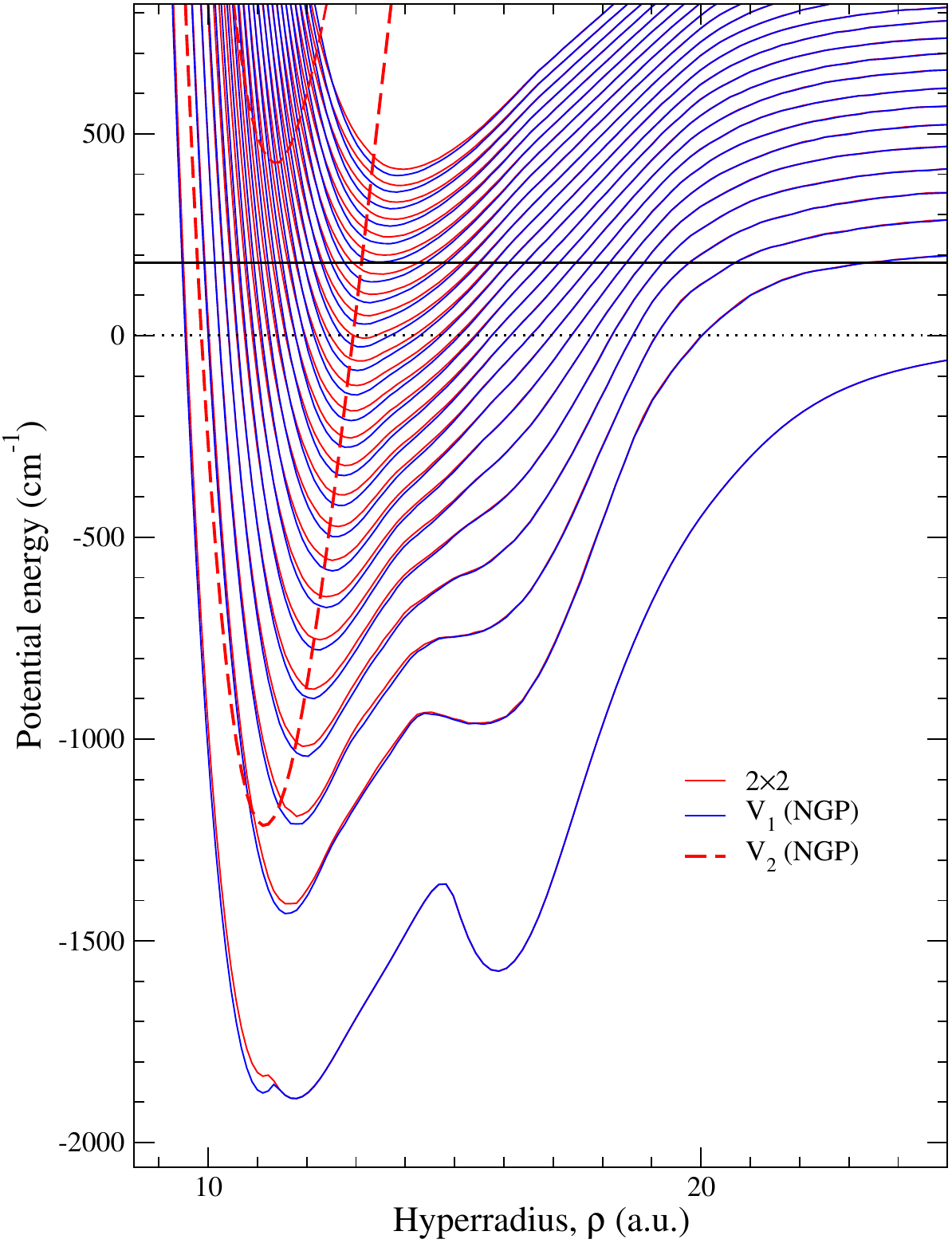}
\par\end{centering}
\caption{\label{fig:NGPvs2x2CoupledCurves}A subset of the lowest APH effective coupled potential curves (cm$^{-1}$) as functions of the hyperradius,
$\rho$ (\textit{a.u.}), where only every 100 curves are displayed. Blue solid lines are used for the $V_{1}$ NGP  curves, red dashed lines for the $V_{2}$ NGP  curves,
and red solid lines for the 2$\times$2 case. The black solid line denotes the KRb$\left(X^{1}\Sigma^{+},\upsilon=0,j=0\right)$
entrance channel energy. All curves are computed
for zero total angular momentum and even exchange symmetry.}
\end{figure}

Figure (\ref{fig:NGPvs2x2CoupledCurves}) presents every hundred $V_{1}$ effective coupled potential curves for the atom-diatom motion within the APH region (in blue)
compared to the first and the hundredth corresponding $V_{2}$ curves (red dashed lines). Both sets of curves are from the NGP formalism. The black
solid line denotes the KRb$\left(X^{1}\Sigma^{+},\upsilon=0,j=0\right)$ entrance channel energy as used in the experiments. Relative to this threshold, the closed
collisional channels from $V_{2}$ are predominantly constrained to the very short-range part of the potential, in the vicinity of the CI
between both electronic states. These are associated
with asymptotic rovibrational states of the excited KRb$\left(a^{3}\Sigma^{+},\upsilon,j\right)$
and K$_{2}\left(a^{3}\Sigma_{u}^{+},\upsilon,j\right)$ molecules,
which are energetically forbidden in collisions of K + KRb$\left(X^{1}\Sigma^{+},\upsilon=0,j=0\right)$
at a few Kelvin of collision energy. For numerical efficiency, the non-adiabatic formalism is only applied
to this interval where the excited state is strongly coupled to the ground state. The extent of this range
is determined by additional convergence studies. The equivalent effective 2$\times$2 potential is depicted with red solid lines in Fig. (\ref{fig:NGPvs2x2CoupledCurves}) and,
as expected, the only noticeable difference between the NGP and 2$\times$2 curves happens to manifest in the vicinity of
the CI. The two sets of curves converge quickly and are indistinguishable well before
$\rho=20$ \textit{a.u.} Thus, beyond this point, a diabatic-to-adiabatic transformation (DAT) is performed on the 2$\times$2 solutions at $\rho_{\mathrm{dat}}$ (see Kendrick~\cite{kendrick2018a} for details), and
the propagation is continued as a usual Born-Oppenheimer calculation. Here, the value of $\rho_{\mathrm{dat}}=30$ \textit{a.u.}
has been chosen as a reasonable compromise between numerical accuracy and computational workload.

Using the NGP  and 2$\times$2 formalisms outlined above, we have evaluated the collision-induced scattering matrix as well as
the state-to-state and total rate coefficients for the KRb$\left(X^{1}\Sigma^{+},\upsilon=0,j=0\right)$ + K $\longrightarrow$
 Rb + K$_{2}\left(X^{1}\Sigma_{g}^{+},\upsilon^{\prime},j^{\prime}\right)$
reaction in the energy range of 2 nK to 0.5 K. The calculations
are made for each independent parity (ortho and para) of the rotational
states of the product K$_{2}$ molecule. The parity-dependent rates
presented below are post-symmetrized in order to take into account
the unresolved nuclear spin degeneracies of the $^{40}$K isotope
as a composite fermion with nuclear spin $I=4$ -- for more details,
see Eq. (39$^{\prime}$) and associated discussion by Miller\cite{miller1969}.
Thus, the total rates from the even and odd exchange symmetry calculations are subsequently
scaled by $\nicefrac{5}{9}$ and $\nicefrac{4}{9}$ factors, respectively.

As a first step, we have attempted to reproduce the original
results of Croft \textit{et al.} \cite{croft2017a,croft2017b}
using the same conventional NGP method within the Born-Oppenheimer
approximation. Here, all grid-dependent parameters are kept nearly
identical to those reported by Croft \textit{et al.} whereas the number
of APH channels is slightly smaller: 2450 (450) channels are deployed
in the APH (Delves) region with respect to 2543 (450) channels used
in the original calculation\cite{croft2017a,croft2017b}. However, this difference
is only expected to have a marginal effect on the total rate constant
of such fairly converged calculations, in particular at higher collision
energies. At 250 nK, we have obtained a total rate constant of $0.9\times10^{-10}$
cm$^{3}$/s, in good agreement with the original result, \textit{i.e.},
$1.1\times10^{-10}$ cm$^{3}$/s \cite{croft2017a,croft2017b}. Part of this difference
can be attributed to minor differences in the adiabatic potential V$_{1}$ adopted in
the work of Croft \textit{et al.}~\cite{croft2017a} compared to that from the present
work, arising from slightly different KRb and K$_{2}$ diatomic potential curves (PECs)
employed in the construction of the PESs utilized here.

Following the same methodology, we invoke the 2$\times$2 method to more accurately describe the dynamics
and to estimate to what extent the NGP  result, and the theory-experiment
disagreement, can be improved as we include the excited electronic
state during the collision. We have propagated 2$\times$2 solutions with 3400 channels from
$\rho_{\mathrm{min}}=8$ \textit{a.u.} to $\rho=25$ \textit{a.u.}, at which point they are projected
onto a smaller subset of the lowest 2543 channels. Then, the propagation is continued
until $\rho_{\mathrm{dat}}=30$ \textit{a.u.}, where the DAT step
is applied onto 2450 NGP solutions for the long-range part of the propagation. This methodology was repeated for choices of $\rho_{\mathrm{match}}=38.1$,
49.9, and 60.3 \textit{a.u.}, where the latter was found to be optimal.

\begin{figure}[H]
\begin{centering}
\includegraphics[scale=0.4]{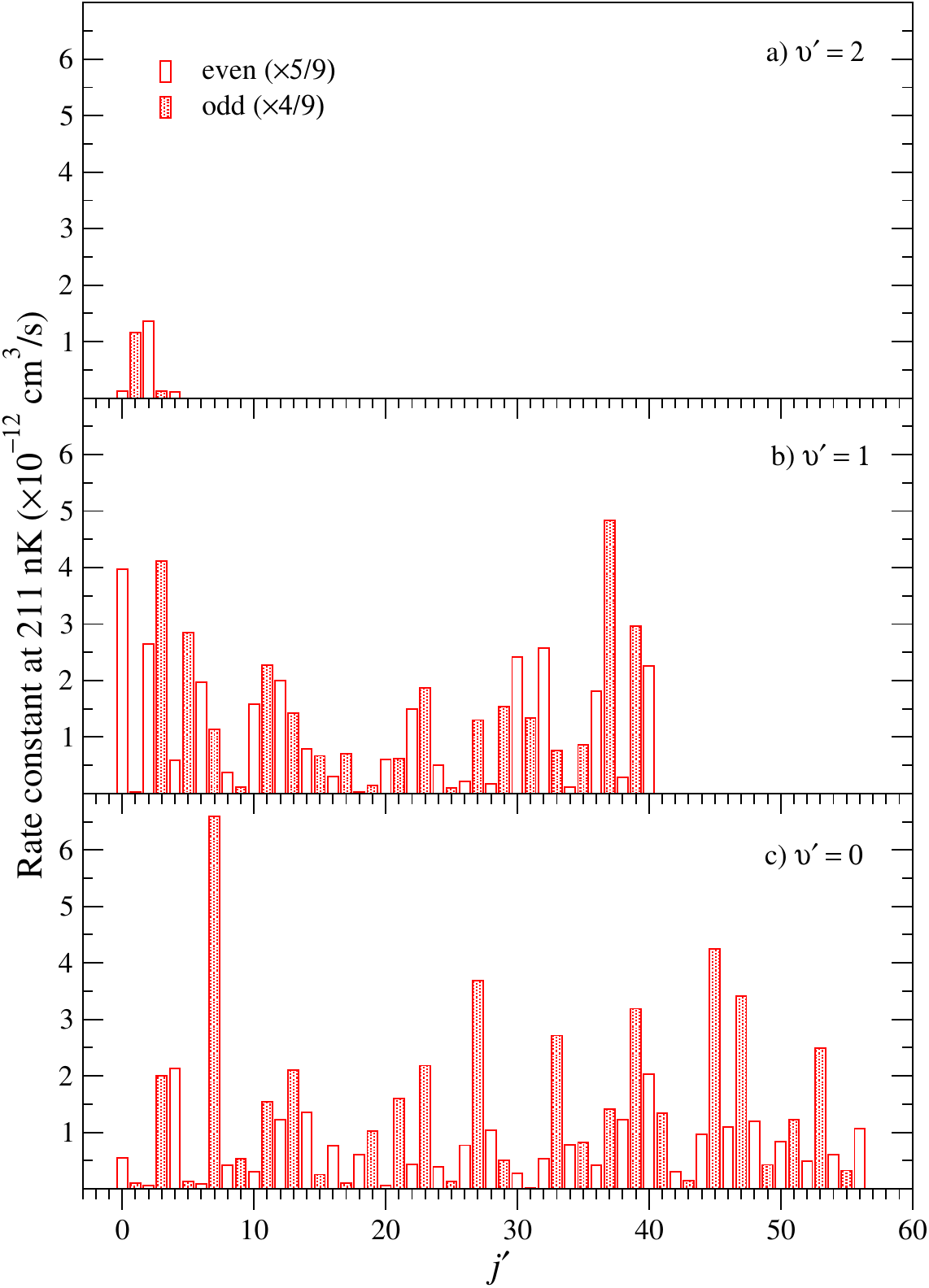}
\par\end{centering}
\caption{\label{fig:RotDistribution} State-to-state rate constants (cm$^{3}$/s)
for the K + KRb$\left(X^{1}\Sigma^{+},\upsilon=0,j=0\right)\protect\longrightarrow$
Rb + K$_{2}\left(X^{1}\Sigma_{g}^{+},\upsilon^{\prime},j^{\prime}\right)$
collision at 211 nK as functions of $j^{\prime}$. Panels (a)--(c)
corresponds to $\upsilon^{\prime}=2$, 1, and 0, respectively. The values of
$\rho_{\mathrm{match}}=60$ \textit{a.u.} and $\rho_{\mathrm{max}}=325$ \textit{a.u.} are used.}
\end{figure}

Rotationally resolved rate coefficients from the 2$\times$2 calculations at 211 nK are
presented in Fig. (\ref{fig:RotDistribution}). The population in each vibrational level
is somewhat similar to that found in the original work of Croft and co-workers at 210 nK\cite{croft2017b}.
Thus, it does confirm a collision-induced formation of highly rotationally excited K$_{2}\left(X^{1}\Sigma_{g}^{+}\right)$
molecules favoring the $\upsilon^{\prime}=0$
and 1 vibrational manifolds. The asymptotic diatomic rovibrational
states used here include all states below a 0.3 eV threshold measured
from the bottom of the ground state PES, \textit{i.e.}, $\upsilon_{\mathrm{max}}=5$,
$j_{\mathrm{max}}=94\:(\upsilon=0)$ for K$_{2}\left(X^{1}\Sigma_{g}^{+}\right)$
and $\upsilon_{\mathrm{max}}=4$, $j_{\mathrm{max}}=90\:(\upsilon=0)$
for the KRb$\left(X^{1}\Sigma^{+}\right)$ molecule. For the typical
collision energies of interest here, only about 55 open channels are
available and populated by the collision.

\begin{figure}[H]
\begin{centering}
\includegraphics[scale=0.4]{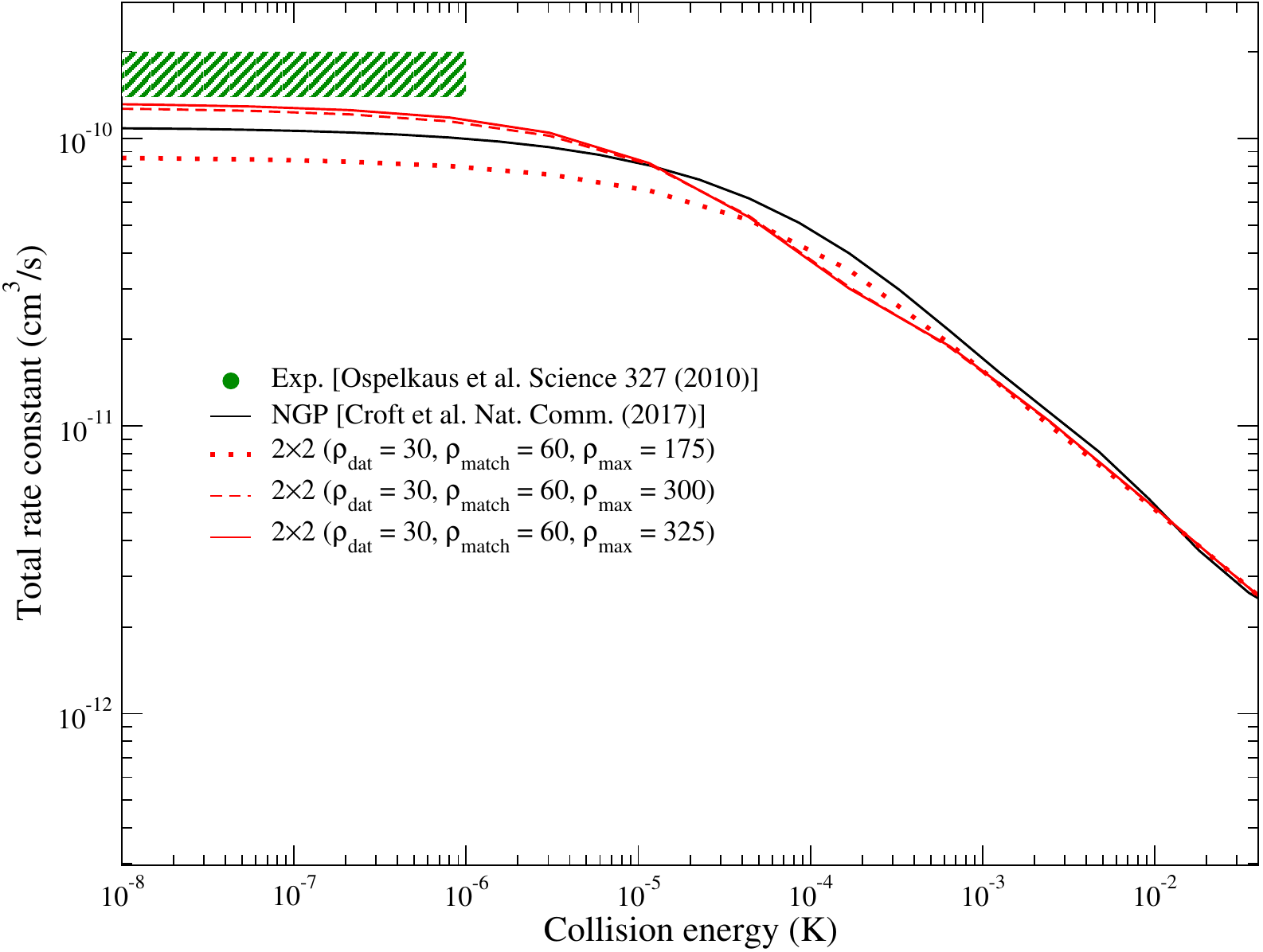}
\par\end{centering}
\caption{\label{fig:NGPvs2x2}Total rate constants (cm$^{3}$/s) as functions
of the collision energy (K). Red curves are for the 2$\times$2 results computed at the displayed values of
$\rho_{\mathrm{dat}}$, $\rho_{\mathrm{match}}$, and $\rho_{\mathrm{max}}$, all in \textit{a.u.}
 The green stripe tags the experimental value of Ospelkaus \textit{et al.}\cite{ospelkaus2010a} within its error bar and uncertain collision energy (below 1 $\mu$K). The black curve is the NGP calculation of Croft \textit{et al.}\cite{croft2017b}.}
\end{figure}

Figure (\ref{fig:NGPvs2x2}) presents the total 2$\times$2 rate constant (in red) and how it fares
in comparison to the original NGP calculation\cite{croft2017b} (in black) and
experiment\cite{ospelkaus2010a} (in green). It is worthwhile to emphasize that the actual temperature
at which the single experimental value has been measured is only known to be below 1 $\mu$K\cite{ospelkaus2010a}.
Here, we do refer to the measurement at 250 nK for the sake of illustration.
Moreover, as evinced in
recent studies\cite{silva2023,morita2023}, scattering characteristics
at the ultra-low range of kinetic energies are notoriously sensitive
to the choice of $\rho_{\mathrm{max}}$, which represents the furthermost
point upon which the Schr$\ddot{\mathrm{o}}$dinger equation is integrated,
and scattering boundary conditions are applied. This is due in part to the unusually large de Broglie
wavelength of the colliding partners as the collision energy is reduced  below 10 $\mu K$. Figure (\ref{fig:NGPvs2x2}) illustrates how the 2$\times$2
and the previous NGP calculations yield nearly indistinguishable results for collision energies greater than 1 mK, regardless of the choice of $\rho_{\mathrm{max}}$ in the  2$\times$2 calculations. A value of $\rho_{\mathrm{max}}=325$ \textit{a.u.} is found to yield nearly identical results (solid red curve in Fig. (\ref{fig:NGPvs2x2})) as $\rho_{\mathrm{max}}=300$ \textit{a.u.} (dashed red curve in Fig. (\ref{fig:NGPvs2x2})) and is taken to be the reference value in our calculations.  This value of $\rho_{\mathrm{max}}$ is
about twice as large as the value used in the original NGP calculation\cite{croft2017b}.

As expected, an inspection of Fig. (\ref{fig:NGPvs2x2}) clearly shows that non-adiabatic quantum interference effects are only
relevant in the ultracold domain, namely below a millikelvin. 
In particular, in the limit where the temperature
approaches zero, the total 2$\times$2 rate constant converges to about 1.32$\times10^{-10}$ cm$^{3}$/s, close to the lower limit of the experimental error bars. As for the actual collision-induced mechanism that is unique to a
chemical reaction subjected to a CI, it has already been addressed
on many occasions\cite{kendrick1996,kendrick2000,kendrick2003b,kendrick2015,kendrick2016a,kendrick2016b}.
Briefly, the scattering wavefunction becomes the sum of two components,
one describing the direct pathway connecting reactants to products
and a second term that accounts for the path looping around the CI.
As a consequence, two incoherent contributions (direct- and loop-dependent)
and one coherent (cross-term) are embedded in the final scattering
amplitude. The coherent term accounts for all constructive and destructive
interference patterns between the two paths. While such an interference effect is
also present in the NGP calculations, the effect of the GP (which is captured in the non-adiabatic treatment)
is to change the sign of this interference term, making constructive to destructive and vice versa.

 It is worth mentioning that while the present calculation improves significantly upon the original work of Croft and co-workers, it does not yet account for
the hyperfine structure within the entrance channel due to the non-null nuclear spins of both K and Rb atoms. In particular, recent experimental studies on the K + NaK\cite{voges2022}
and Rb + KRb\cite{liu2025} systems have found strong evidence of hyperfine-dependent collisional outcomes at fixed magnetic fields that cannot be neglected. However, solving numerically an atom-diatom
Hamiltonian that accounts for the hyperfine structure in addition to collision-induced rearrangements and non-adiabatic effects in
hyperspherical coordinates is no trivial task and beyond the scope of this work. Yet our results are relevant in light of a number of similar discrepancies between theory and experiments reported
thus far. In particular, near-universal losses have been consistently observed over the years
in experiments on ultracold mixtures of such dipolar bialkali dimers, even if chemically
stable ground-state molecules are utilized\cite{takekoshi2014,park2015,voges2020,guo2018,ye2018,gregory2019,gersema2021,bause2021}.
Likewise, the original work of Ospelkaus \textit{et al.} \cite{ospelkaus2010a} also measured the loss rate of
KRb molecules in the presence of Rb atoms, an ultracold mixture whose
reactive process is endothermic by about 214 cm$^{-1}$.
Yet the number of KRb molecules is observed to decay by a rate
about one order of magnitude slower than that of K + KRb\cite{ospelkaus2010a}.
This observation has been confirmed recently by Nichols \textit{et
al.}, finding further evidence that suggests the formation of unusually
long-lived triatomic complexes during the collision\cite{nichols2022,bause2023}.
Such a discrepancy may have also been indirectly observed in recent
experiments on NaRb and NaK\cite{gersema2021,bause2021}.

The predominant explanation has been the sticky collisions hypothesis
proposed by Mayle \textit{et al.}\cite{Mayle2013} which highly depends
on one's ability to accurately model the collision-induced formation
of the complex and its subsequent evolution. However, a proper characterization
of the two-body loss within such mixtures has been considered
incomplete at best\cite{bause2023}, as the lifetimes of the collision-induced complex
have been observed to deviate by orders of magnitude from recent predictions.

Theoretical models used to describe the collision-induced formation of the complex and the observed
two-body losses rely mostly on pure long-range interactions, followed by oftentimes statistical descriptions of the dynamics.
While these methodologies have been seen as valid in certain conditions, the unique non-adiabatically driven quantum effects on the title
reaction, as shown here, do illustrate the need for a rigorous treatment of the short-range interaction. Within the validity of the PESs utilized in the present work,
an improved agreement with experiments was achieved by means of such a treatment. In particular, the properties
of the short-range atom-diatom interaction mediated by coupling with the excited electronic state play a central role and cannot
be neglected. Our findings may hint at the direction in which most of the discrepancies noted above could be addressed and perhaps explained.

\acknowledgement

This work is supported in part by NSF grants PHY-2110227 \& PHY-2409497 (N.B.) and a DoD MURI grant from Army Research Office
(Grant No. W911NF-19-1-0283) (N.B.). It is also supported (S.K. and H.I.) by NSF grant PHY-2409425 and AFORS grant FA9550-21-1-0153.
B.K.K. acknowledges that part of this work was done under the auspices of the US Department of Energy under Project No. 20240256ER
of the Laboratory Directed Research and Development Program at Los Alamos National Laboratory. This work used resources provided by
the Los Alamos National Laboratory Institutional Computing Program. Los Alamos National Laboratory is operated by Triad National
Security, LLC, for the National Nuclear Security Administration of the U.S. Department of Energy (contract No. 89233218CNA000001).
This work used Bridges-2 resources at Pittsburgh Supercomputing Center (PSC) through allocation PHY200034 (N.B.) from the Advanced Cyberinfrastructure Coordination Ecosystem: Services \& Support (ACCESS) program, which is supported by U.S. National Science Foundation grants \#2138259, \#2138286, \#2138307, \#2137603, and \#2138296.

\bibliography{references}

\providecommand{\latin}[1]{#1}
\makeatletter
\providecommand{\doi}
  {\begingroup\let\do\@makeother\dospecials
  \catcode`\{=1 \catcode`\}=2 \doi@aux}
\providecommand{\doi@aux}[1]{\endgroup\texttt{#1}}
\makeatother
\providecommand*\mcitethebibliography{\thebibliography}
\csname @ifundefined\endcsname{endmcitethebibliography}
  {\let\endmcitethebibliography\endthebibliography}{}
\begin{mcitethebibliography}{65}
\providecommand*\natexlab[1]{#1}
\providecommand*\mciteSetBstSublistMode[1]{}
\providecommand*\mciteSetBstMaxWidthForm[2]{}
\providecommand*\mciteBstWouldAddEndPuncttrue
  {\def\EndOfBibitem{\unskip.}}
\providecommand*\mciteBstWouldAddEndPunctfalse
  {\let\EndOfBibitem\relax}
\providecommand*\mciteSetBstMidEndSepPunct[3]{}
\providecommand*\mciteSetBstSublistLabelBeginEnd[3]{}
\providecommand*\EndOfBibitem{}
\mciteSetBstSublistMode{f}
\mciteSetBstMaxWidthForm{subitem}{(\alph{mcitesubitemcount})}
\mciteSetBstSublistLabelBeginEnd
  {\mcitemaxwidthsubitemform\space}
  {\relax}
  {\relax}

\bibitem[Krems(2008)]{krems2008}
Krems,~R.~V. Cold controlled chemistry. \emph{Phys. Chem. Chem. Phys.}
  \textbf{2008}, \emph{10}, 4079--4092\relax
\mciteBstWouldAddEndPuncttrue
\mciteSetBstMidEndSepPunct{\mcitedefaultmidpunct}
{\mcitedefaultendpunct}{\mcitedefaultseppunct}\relax
\EndOfBibitem
\bibitem[Carr and Ye(2009)Carr, and Ye]{Carr2009}
Carr,~L.~D.; Ye,~J. Focus on Cold and Ultracold Molecules. \emph{New J. Phys.}
  \textbf{2009}, \emph{11}, 055009\relax
\mciteBstWouldAddEndPuncttrue
\mciteSetBstMidEndSepPunct{\mcitedefaultmidpunct}
{\mcitedefaultendpunct}{\mcitedefaultseppunct}\relax
\EndOfBibitem
\bibitem[Kendrick \latin{et~al.}(2015)Kendrick, Hazra, and
  Balakrishnan]{kendrick2015}
Kendrick,~B.~K.; Hazra,~J.; Balakrishnan,~N. The geometric phase controls
  ultracold chemistry. \emph{Nature Communications} \textbf{2015}, \emph{6},
  7918\relax
\mciteBstWouldAddEndPuncttrue
\mciteSetBstMidEndSepPunct{\mcitedefaultmidpunct}
{\mcitedefaultendpunct}{\mcitedefaultseppunct}\relax
\EndOfBibitem
\bibitem[Balakrishnan(2016)]{balakrishnan2016}
Balakrishnan,~N. Perspective: Ultracold molecules and the dawn of cold
  controlled chemistry. \emph{The Journal of Chemical Physics} \textbf{2016},
  \emph{145}\relax
\mciteBstWouldAddEndPuncttrue
\mciteSetBstMidEndSepPunct{\mcitedefaultmidpunct}
{\mcitedefaultendpunct}{\mcitedefaultseppunct}\relax
\EndOfBibitem
\bibitem[Bohn \latin{et~al.}(2017)Bohn, Rey, and Ye]{Bohn2017}
Bohn,~J.~L.; Rey,~A.~M.; Ye,~J. Cold molecules: Progress in quantum engineering
  of chemistry and quantum matter. \emph{Science} \textbf{2017}, \emph{357},
  1002\relax
\mciteBstWouldAddEndPuncttrue
\mciteSetBstMidEndSepPunct{\mcitedefaultmidpunct}
{\mcitedefaultendpunct}{\mcitedefaultseppunct}\relax
\EndOfBibitem
\bibitem[Hu \latin{et~al.}(2019)Hu, Liu, Grimes, Lin, Gheorghe, Vexiau,
  Bouloufa-Maafa, Dulieu, Rosenband, and Ni]{hu2019}
Hu,~M.-G.; Liu,~Y.; Grimes,~D.~D.; Lin,~Y.-W.; Gheorghe,~A.~H.; Vexiau,~R.;
  Bouloufa-Maafa,~N.; Dulieu,~O.; Rosenband,~T.; Ni,~K.-K. Direct observation
  of bimolecular reactions of ultracold {KRb} molecules. \emph{Science}
  \textbf{2019}, \emph{366}, 1111--1115\relax
\mciteBstWouldAddEndPuncttrue
\mciteSetBstMidEndSepPunct{\mcitedefaultmidpunct}
{\mcitedefaultendpunct}{\mcitedefaultseppunct}\relax
\EndOfBibitem
\bibitem[Liu \latin{et~al.}(2021)Liu, Hu, Nichols, Yang, Xie, Guo, and
  Ni]{Liu2021}
Liu,~Y.; Hu,~M.-G.; Nichols,~M.~A.; Yang,~D.; Xie,~D.; Guo,~H.; Ni,~K.-K.
  Precision test of statistical dynamics with state-to-state ultracold
  chemistry. \emph{Science} \textbf{2021}, \emph{593}, 379--384\relax
\mciteBstWouldAddEndPuncttrue
\mciteSetBstMidEndSepPunct{\mcitedefaultmidpunct}
{\mcitedefaultendpunct}{\mcitedefaultseppunct}\relax
\EndOfBibitem
\bibitem[Kendrick \latin{et~al.}(2021)Kendrick, Li, Li, Kotochigova, Croft, and
  Balakrishnan]{kendrick2021}
Kendrick,~B.~K.; Li,~H.; Li,~M.; Kotochigova,~S.; Croft,~J. F.~E.;
  Balakrishnan,~N. Non-adiabatic quantum interference in the ultracold {Li +
  LiNa} $\rightarrow$ {Li$_{2}$ + Na} reaction. \emph{Phys. Chem. Chem. Phys.}
  \textbf{2021}, \emph{23}, 5096--5112\relax
\mciteBstWouldAddEndPuncttrue
\mciteSetBstMidEndSepPunct{\mcitedefaultmidpunct}
{\mcitedefaultendpunct}{\mcitedefaultseppunct}\relax
\EndOfBibitem
\bibitem[Liu and Ni(2022)Liu, and Ni]{Liu2022}
Liu,~Y.; Ni,~K.~K. Bimolecular chemistry in the ultracold regime. \emph{Annual
  Review of Physical Chemistry} \textbf{2022}, \emph{73}, 73\relax
\mciteBstWouldAddEndPuncttrue
\mciteSetBstMidEndSepPunct{\mcitedefaultmidpunct}
{\mcitedefaultendpunct}{\mcitedefaultseppunct}\relax
\EndOfBibitem
\bibitem[Kendrick \latin{et~al.}(2015)Kendrick, Hazra, and
  Balakrishnan]{Kendrick_2015_PRL}
Kendrick,~B.~K.; Hazra,~J.; Balakrishnan,~N. Geometric Phase Appears in the
  Ultracold Hydrogen Exchange Reaction. \emph{Phys. Rev. Lett.} \textbf{2015},
  \emph{115}, 153201\relax
\mciteBstWouldAddEndPuncttrue
\mciteSetBstMidEndSepPunct{\mcitedefaultmidpunct}
{\mcitedefaultendpunct}{\mcitedefaultseppunct}\relax
\EndOfBibitem
\bibitem[Hazra \latin{et~al.}(2016)Hazra, Kendrick, and
  Balakrishnan]{hazra2016}
Hazra,~J.; Kendrick,~B.~K.; Balakrishnan,~N. Geometric phase effects in
  ultracold hydrogen exchange reaction. \emph{Journal of Physics B: Atomic,
  Molecular and Optical Physics} \textbf{2016}, \emph{49}, 194004\relax
\mciteBstWouldAddEndPuncttrue
\mciteSetBstMidEndSepPunct{\mcitedefaultmidpunct}
{\mcitedefaultendpunct}{\mcitedefaultseppunct}\relax
\EndOfBibitem
\bibitem[Kendrick \latin{et~al.}(2016)Kendrick, Hazra, and
  Balakrishnan]{kendrick2016a}
Kendrick,~B.~K.; Hazra,~J.; Balakrishnan,~N. Geometric phase effects in the
  ultracold {H + H$_{2}$} reaction. \emph{The Journal of Chemical Physics}
  \textbf{2016}, \emph{145}, 164303\relax
\mciteBstWouldAddEndPuncttrue
\mciteSetBstMidEndSepPunct{\mcitedefaultmidpunct}
{\mcitedefaultendpunct}{\mcitedefaultseppunct}\relax
\EndOfBibitem
\bibitem[Kendrick \latin{et~al.}(2016)Kendrick, Hazra, and
  Balakrishnan]{kendrick2016b}
Kendrick,~B.~K.; Hazra,~J.; Balakrishnan,~N. Geometric phase effects in the
  ultracold {D + HD $\rightarrow$ D + HD} and {D + HD $\leftrightarrow$ H +
  D$_{2}$} reactions. \emph{New Journal of Physics} \textbf{2016}, \emph{18},
  123020\relax
\mciteBstWouldAddEndPuncttrue
\mciteSetBstMidEndSepPunct{\mcitedefaultmidpunct}
{\mcitedefaultendpunct}{\mcitedefaultseppunct}\relax
\EndOfBibitem
\bibitem[Kendrick(2018)]{kendrick2018b}
Kendrick,~B.~K. Non-adiabatic quantum reactive scattering calculations for the
  ultracold hydrogen exchange reaction: {H}+{H}$_2(v=4-8,j=0)\to$
  {H}+{H}$_2(v',j')$. \emph{Chemical Physics} \textbf{2018}, \emph{515},
  387--399\relax
\mciteBstWouldAddEndPuncttrue
\mciteSetBstMidEndSepPunct{\mcitedefaultmidpunct}
{\mcitedefaultendpunct}{\mcitedefaultseppunct}\relax
\EndOfBibitem
\bibitem[Kendrick(2019)]{kendrick2019}
Kendrick,~B.~K. Nonadiabatic Ultracold Quantum Reactive Scattering of Hydrogen
  with Vibrationally Excited {HD}($v = 5-9$). \emph{The Journal of Physical
  Chemistry A} \textbf{2019}, \emph{123}, 9919--9933\relax
\mciteBstWouldAddEndPuncttrue
\mciteSetBstMidEndSepPunct{\mcitedefaultmidpunct}
{\mcitedefaultendpunct}{\mcitedefaultseppunct}\relax
\EndOfBibitem
\bibitem[Hazra \latin{et~al.}(2015)Hazra, Kendrick, and
  Balakrishnan]{Hazra_JPCA_2015}
Hazra,~J.; Kendrick,~B.~K.; Balakrishnan,~N. Importance of Geometric Phase
  Effects in Ultracold Chemistry. \emph{The Journal of Physical Chemistry A}
  \textbf{2015}, \emph{119}, 12291--12303\relax
\mciteBstWouldAddEndPuncttrue
\mciteSetBstMidEndSepPunct{\mcitedefaultmidpunct}
{\mcitedefaultendpunct}{\mcitedefaultseppunct}\relax
\EndOfBibitem
\bibitem[Wang \latin{et~al.}(2025)Wang, Hu, Guo, and Xie]{wang2025}
Wang,~J.; Hu,~X.; Guo,~H.; Xie,~D. Cold {H + O$_{2}$} collisions: Impact of
  resonances, geometric phase, and alignment. \emph{The Journal of Chemical
  Physics} \textbf{2025}, \emph{162}, 074301\relax
\mciteBstWouldAddEndPuncttrue
\mciteSetBstMidEndSepPunct{\mcitedefaultmidpunct}
{\mcitedefaultendpunct}{\mcitedefaultseppunct}\relax
\EndOfBibitem
\bibitem[Ospelkaus \latin{et~al.}(2010)Ospelkaus, Ni, Wang, de~Miranda,
  Neyenhuis, Qu\'em\'ener, Julienne, Bohn, Jin, and Ye]{ospelkaus2010a}
Ospelkaus,~S.; Ni,~K.-K.; Wang,~D.; de~Miranda,~M. H.~G.; Neyenhuis,~B.;
  Qu\'em\'ener,~G.; Julienne,~P.~S.; Bohn,~J.; Jin,~D.~S.; Ye,~J. Quantum state
  controlled chemical reactions of ultracold potassium-rubidium molecules.
  \emph{Science} \textbf{2010}, \emph{327}, 853\relax
\mciteBstWouldAddEndPuncttrue
\mciteSetBstMidEndSepPunct{\mcitedefaultmidpunct}
{\mcitedefaultendpunct}{\mcitedefaultseppunct}\relax
\EndOfBibitem
\bibitem[Liu \latin{et~al.}(2024)Liu, Zhu, Luke, Houwman, Babin, Hu, and
  Ni]{Liu_KRb_2024}
Liu,~Y.~X.; Zhu,~L.; Luke,~J.; Houwman,~J. J.~A.; Babin,~M.~C.; Hu,~M.~G.;
  Ni,~K.~K. Quantum interference in atom-exchange reactions. \emph{Science}
  \textbf{2024}, \emph{384}, 1117\relax
\mciteBstWouldAddEndPuncttrue
\mciteSetBstMidEndSepPunct{\mcitedefaultmidpunct}
{\mcitedefaultendpunct}{\mcitedefaultseppunct}\relax
\EndOfBibitem
\bibitem[Croft \latin{et~al.}(2017)Croft, Balakrishnan, and
  Kendrick]{croft2017a}
Croft,~J. F.~E.; Balakrishnan,~N.; Kendrick,~B.~K. Long-lived complexes and
  signatures of chaos in ultracold ${\mathrm{K}}_{2}${+Rb} collisions.
  \emph{Phys. Rev. A} \textbf{2017}, \emph{96}, 062707\relax
\mciteBstWouldAddEndPuncttrue
\mciteSetBstMidEndSepPunct{\mcitedefaultmidpunct}
{\mcitedefaultendpunct}{\mcitedefaultseppunct}\relax
\EndOfBibitem
\bibitem[Yarkony(1996)]{yarkony1996}
Yarkony,~D.~R. Diabolical conical intersections. \emph{Rev. Mod. Phys.}
  \textbf{1996}, \emph{68}, 985--1013\relax
\mciteBstWouldAddEndPuncttrue
\mciteSetBstMidEndSepPunct{\mcitedefaultmidpunct}
{\mcitedefaultendpunct}{\mcitedefaultseppunct}\relax
\EndOfBibitem
\bibitem[Domcke and Yarkony(2012)Domcke, and Yarkony]{domcke2012}
Domcke,~W.; Yarkony,~D.~R. Role of Conical Intersections in Molecular
  Spectroscopy and Photoinduced Chemical Dynamics. \emph{Annual Review of
  Physical Chemistry} \textbf{2012}, \emph{63}, 325--352, PMID: 22475338\relax
\mciteBstWouldAddEndPuncttrue
\mciteSetBstMidEndSepPunct{\mcitedefaultmidpunct}
{\mcitedefaultendpunct}{\mcitedefaultseppunct}\relax
\EndOfBibitem
\bibitem[Tully(2012)]{tully2012}
Tully,~J.~C. Perspective: Nonadiabatic dynamics theory. \emph{The Journal of
  Chemical Physics} \textbf{2012}, \emph{137}, 22A301\relax
\mciteBstWouldAddEndPuncttrue
\mciteSetBstMidEndSepPunct{\mcitedefaultmidpunct}
{\mcitedefaultendpunct}{\mcitedefaultseppunct}\relax
\EndOfBibitem
\bibitem[Guo and Yarkony(2016)Guo, and Yarkony]{guo2016}
Guo,~H.; Yarkony,~D.~R. Accurate nonadiabatic dynamics. \emph{Phys. Chem. Chem.
  Phys.} \textbf{2016}, \emph{18}, 26335--26352\relax
\mciteBstWouldAddEndPuncttrue
\mciteSetBstMidEndSepPunct{\mcitedefaultmidpunct}
{\mcitedefaultendpunct}{\mcitedefaultseppunct}\relax
\EndOfBibitem
\bibitem[Huang \latin{et~al.}(2018)Huang, Liu, Zhang, and Krems]{huang2018}
Huang,~J.; Liu,~S.; Zhang,~D.~H.; Krems,~R.~V. Time-Dependent Wave Packet
  Dynamics Calculations of Cross Sections for Ultracold Scattering of
  Molecules. \emph{Phys. Rev. Lett.} \textbf{2018}, \emph{120}, 143401\relax
\mciteBstWouldAddEndPuncttrue
\mciteSetBstMidEndSepPunct{\mcitedefaultmidpunct}
{\mcitedefaultendpunct}{\mcitedefaultseppunct}\relax
\EndOfBibitem
\bibitem[Kendrick(2018)]{kendrick2018a}
Kendrick,~B.~K. Non-adiabatic quantum reactive scattering in hyperspherical
  coordinates. \emph{The Journal of Chemical Physics} \textbf{2018},
  \emph{148}, 044116\relax
\mciteBstWouldAddEndPuncttrue
\mciteSetBstMidEndSepPunct{\mcitedefaultmidpunct}
{\mcitedefaultendpunct}{\mcitedefaultseppunct}\relax
\EndOfBibitem
\bibitem[Longuet-Higgins \latin{et~al.}(1958)Longuet-Higgins, Opik, Pryce, and
  Sack]{higgins1958}
Longuet-Higgins,~H.~C.; Opik,~U.; Pryce,~M. H.~L.; Sack,~R.~A. Studies of the
  {Jahn-Teller} effect .{II.} The dynamical problem. \emph{Proceedings of the
  Royal Society of London. Series A. Mathematical and Physical Sciences}
  \textbf{1958}, \emph{244}, 1--16\relax
\mciteBstWouldAddEndPuncttrue
\mciteSetBstMidEndSepPunct{\mcitedefaultmidpunct}
{\mcitedefaultendpunct}{\mcitedefaultseppunct}\relax
\EndOfBibitem
\bibitem[Herzberg and Longuet-Higgins(1963)Herzberg, and
  Longuet-Higgins]{herzberg1963}
Herzberg,~G.; Longuet-Higgins,~H.~C. Intersection of potential energy surfaces
  in polyatomic molecules. \emph{Discuss. Faraday Soc.} \textbf{1963},
  \emph{35}, 77--82\relax
\mciteBstWouldAddEndPuncttrue
\mciteSetBstMidEndSepPunct{\mcitedefaultmidpunct}
{\mcitedefaultendpunct}{\mcitedefaultseppunct}\relax
\EndOfBibitem
\bibitem[O'Brien and Bleaney(1964)O'Brien, and Bleaney]{obrien1964}
O'Brien,~M. C.~M.; Bleaney,~B. The dynamic {Jahn-Teller} effect in octahedrally
  co-ordinated $d^{9}$ ions. \emph{Proceedings of the Royal Society of London.
  Series A. Mathematical and Physical Sciences} \textbf{1964}, \emph{281},
  323--339\relax
\mciteBstWouldAddEndPuncttrue
\mciteSetBstMidEndSepPunct{\mcitedefaultmidpunct}
{\mcitedefaultendpunct}{\mcitedefaultseppunct}\relax
\EndOfBibitem
\bibitem[Kendrick(2000)]{kendrick2000}
Kendrick,~B.~K. Geometric phase effects in the {H+D}$_{2}\rightarrow${HD+D}
  reaction. \emph{The Journal of Chemical Physics} \textbf{2000}, \emph{112},
  5679--5704\relax
\mciteBstWouldAddEndPuncttrue
\mciteSetBstMidEndSepPunct{\mcitedefaultmidpunct}
{\mcitedefaultendpunct}{\mcitedefaultseppunct}\relax
\EndOfBibitem
\bibitem[Kendrick(2001)]{kendrick2001}
Kendrick,~B.~K. Quantum reactive scattering calculations for the
  {H+D}$_{2}\rightarrow${HD+D} reaction. \emph{The Journal of Chemical Physics}
  \textbf{2001}, \emph{114}, 8796--8819\relax
\mciteBstWouldAddEndPuncttrue
\mciteSetBstMidEndSepPunct{\mcitedefaultmidpunct}
{\mcitedefaultendpunct}{\mcitedefaultseppunct}\relax
\EndOfBibitem
\bibitem[Kendrick(2003)]{kendrick2003a}
Kendrick,~B.~K. Quantum reactive scattering calculations for the
  {D}+{H}$_{2}\rightarrow${HD}+{H} reaction. \emph{The Journal of Chemical
  Physics} \textbf{2003}, \emph{118}, 10502--10522\relax
\mciteBstWouldAddEndPuncttrue
\mciteSetBstMidEndSepPunct{\mcitedefaultmidpunct}
{\mcitedefaultendpunct}{\mcitedefaultseppunct}\relax
\EndOfBibitem
\bibitem[Kendrick(2003)]{kendrick2003b}
Kendrick,~B.~K. Geometric Phase Effects in Chemical Reaction Dynamics and
  Molecular Spectra. \emph{The Journal of Physical Chemistry A} \textbf{2003},
  \emph{107}, 6739--6756\relax
\mciteBstWouldAddEndPuncttrue
\mciteSetBstMidEndSepPunct{\mcitedefaultmidpunct}
{\mcitedefaultendpunct}{\mcitedefaultseppunct}\relax
\EndOfBibitem
\bibitem[Makrides \latin{et~al.}(2015)Makrides, Hazra, Pradhan, Petrov,
  Kendrick, Gonz\'alez-Lezana, Balakrishnan, and Kotochigova]{makrides2015}
Makrides,~C.; Hazra,~J.; Pradhan,~G.~B.; Petrov,~A.; Kendrick,~B.~K.;
  Gonz\'alez-Lezana,~T.; Balakrishnan,~N.; Kotochigova,~S. Ultracold chemistry
  with alkali-metal--rare-earth molecules. \emph{Phys. Rev. A} \textbf{2015},
  \emph{91}, 012708\relax
\mciteBstWouldAddEndPuncttrue
\mciteSetBstMidEndSepPunct{\mcitedefaultmidpunct}
{\mcitedefaultendpunct}{\mcitedefaultseppunct}\relax
\EndOfBibitem
\bibitem[Croft \latin{et~al.}(2017)Croft, Makrides, Li, Petrov, Kendrick,
  Balakrishnan, and Kotochigova]{croft2017b}
Croft,~J. F.~E.; Makrides,~C.; Li,~M.; Petrov,~A.; Kendrick,~B.~K.;
  Balakrishnan,~N.; Kotochigova,~S. Universality and chaoticity in ultracold
  {K+KRb} chemical reactions. \emph{Nature Communications} \textbf{2017},
  \emph{8}\relax
\mciteBstWouldAddEndPuncttrue
\mciteSetBstMidEndSepPunct{\mcitedefaultmidpunct}
{\mcitedefaultendpunct}{\mcitedefaultseppunct}\relax
\EndOfBibitem
\bibitem[da~Silva \latin{et~al.}(2022)da~Silva, Kendrick, and
  Balakrishnan]{silva2022a}
da~Silva,~H.; Kendrick,~B.~K.; Balakrishnan,~N. On the use of stereodynamical
  effects to control cold chemical reactions: The {H + D}$_{2} \rightarrow$ {D
  + HD} case study. \emph{The Journal of Chemical Physics} \textbf{2022},
  \emph{156}, 044305\relax
\mciteBstWouldAddEndPuncttrue
\mciteSetBstMidEndSepPunct{\mcitedefaultmidpunct}
{\mcitedefaultendpunct}{\mcitedefaultseppunct}\relax
\EndOfBibitem
\bibitem[da~Silva \latin{et~al.}(2023)da~Silva, Yao, Morita, Kendrick, Guo, and
  Balakrishnan]{silva2023}
da~Silva,~H.; Yao,~Q.; Morita,~M.; Kendrick,~B.~K.; Guo,~H.; Balakrishnan,~N.
  The {Li + CaF} $\rightarrow$ {Ca + LiF} chemical reaction under cold
  conditions. \emph{Phys. Chem. Chem. Phys.} \textbf{2023}, \emph{25},
  14193--14205\relax
\mciteBstWouldAddEndPuncttrue
\mciteSetBstMidEndSepPunct{\mcitedefaultmidpunct}
{\mcitedefaultendpunct}{\mcitedefaultseppunct}\relax
\EndOfBibitem
\bibitem[Morita \latin{et~al.}(2023)Morita, Kendrick, Kłos, Kotochigova,
  Brumer, and Tscherbul]{morita2023}
Morita,~M.; Kendrick,~B.~K.; Kłos,~J.; Kotochigova,~S.; Brumer,~P.;
  Tscherbul,~T.~V. Signatures of Non-universal Quantum Dynamics of Ultracold
  Chemical Reactions of Polar Alkali Dimer Molecules with Alkali Metal Atoms:
  {Li($^{2}$S)} + {NaLi}(a$^{3}\Sigma^{+}$) $\rightarrow$ {Na($^{2}$S)} +
  {Li}$_{2}$(a$^{3}\Sigma u^{+}$). \emph{The Journal of Physical Chemistry
  Letters} \textbf{2023}, \emph{14}, 3413--3421, PMID: 37001115\relax
\mciteBstWouldAddEndPuncttrue
\mciteSetBstMidEndSepPunct{\mcitedefaultmidpunct}
{\mcitedefaultendpunct}{\mcitedefaultseppunct}\relax
\EndOfBibitem
\bibitem[Pack and Parker(1987)Pack, and Parker]{pack1987}
Pack,~R.~T.; Parker,~G.~A. Quantum reactive scattering in three dimensions
  using hyperspherical {(APH)} coordinates. {T}heory. \emph{J. Chem. Phys.}
  \textbf{1987}, \emph{87}, 3888\relax
\mciteBstWouldAddEndPuncttrue
\mciteSetBstMidEndSepPunct{\mcitedefaultmidpunct}
{\mcitedefaultendpunct}{\mcitedefaultseppunct}\relax
\EndOfBibitem
\bibitem[Delves(1958)]{delves1958}
Delves,~L.~M. Tertiary and general-order collisions. \emph{Nuclear Physics}
  \textbf{1958}, \emph{9}, 391--399\relax
\mciteBstWouldAddEndPuncttrue
\mciteSetBstMidEndSepPunct{\mcitedefaultmidpunct}
{\mcitedefaultendpunct}{\mcitedefaultseppunct}\relax
\EndOfBibitem
\bibitem[Delves(1960)]{delves1960}
Delves,~L.~M. Tertiary and general-order collisions ({II}). \emph{Nuclear
  Physics} \textbf{1960}, \emph{20}, 275--308\relax
\mciteBstWouldAddEndPuncttrue
\mciteSetBstMidEndSepPunct{\mcitedefaultmidpunct}
{\mcitedefaultendpunct}{\mcitedefaultseppunct}\relax
\EndOfBibitem
\bibitem[Johnson(1973)]{johnson1973}
Johnson,~B. The multichannel log-derivative method for scattering calculations.
  \emph{Journal of Computational Physics} \textbf{1973}, \emph{13},
  445--449\relax
\mciteBstWouldAddEndPuncttrue
\mciteSetBstMidEndSepPunct{\mcitedefaultmidpunct}
{\mcitedefaultendpunct}{\mcitedefaultseppunct}\relax
\EndOfBibitem
\bibitem[Baer(2002)]{Baer2002}
Baer,~M. Introduction to the theory of electronic non-adiabatic coupling terms
  in molecular systems. \emph{Phys. Rep.} \textbf{2002}, \emph{358}, 75 --
  142\relax
\mciteBstWouldAddEndPuncttrue
\mciteSetBstMidEndSepPunct{\mcitedefaultmidpunct}
{\mcitedefaultendpunct}{\mcitedefaultseppunct}\relax
\EndOfBibitem
\bibitem[Fuentealba \latin{et~al.}(1983)Fuentealba, Stoll, von Szentpaly,
  Schwerdtfeger, and Preuss]{Fuentealba1983_1}
Fuentealba,~P.; Stoll,~H.; von Szentpaly,~L.; Schwerdtfeger,~P.; Preuss,~H. On
  the reliability of semi-empirical pseudopotentials: simulation of
  Hartree-Fock and Dirac-Fock results. \emph{Journal of Physics B: Atomic and
  Molecular Physics} \textbf{1983}, \emph{16}, L323\relax
\mciteBstWouldAddEndPuncttrue
\mciteSetBstMidEndSepPunct{\mcitedefaultmidpunct}
{\mcitedefaultendpunct}{\mcitedefaultseppunct}\relax
\EndOfBibitem
\bibitem[Fuentealba \latin{et~al.}(1983)Fuentealba, Szentp{\'a}ly, Stoll,
  Fraschio, and Preuss]{Fuentealba1983_2}
Fuentealba,~P.; Szentp{\'a}ly,~L.; Stoll,~H.; Fraschio,~F.~X.; Preuss,~H.
  Pseudopotential calculations including core-valence correlation: {A}lkali
  compounds. \emph{J. Mol. Struct.: THEOCHEM} \textbf{1983}, \emph{93},
  213--219\relax
\mciteBstWouldAddEndPuncttrue
\mciteSetBstMidEndSepPunct{\mcitedefaultmidpunct}
{\mcitedefaultendpunct}{\mcitedefaultseppunct}\relax
\EndOfBibitem
\bibitem[{ \.Z}uchowski and Hutson(2010){ \.Z}uchowski, and
  Hutson]{Zuchowski2010}
{ \.Z}uchowski,~P.~S.; Hutson,~J.~M. Reactions of ultracold alkali-metal
  dimers. \emph{Phys. Rev. A} \textbf{2010}, \emph{81}, 060703\relax
\mciteBstWouldAddEndPuncttrue
\mciteSetBstMidEndSepPunct{\mcitedefaultmidpunct}
{\mcitedefaultendpunct}{\mcitedefaultseppunct}\relax
\EndOfBibitem
\bibitem[Werner \latin{et~al.}(2012)Werner, Knowles, Knizia, Manby,
  Schu{\"u}tz, Celani, Korona, Lindh, Mitrushenkov, Rauhut, \latin{et~al.}
  others]{Werner2012}
Werner,~H.-J.; Knowles,~P.~J.; Knizia,~G.; Manby,~F.~R.; Schu{\"u}tz,~M.;
  Celani,~P.; Korona,~T.; Lindh,~R.; Mitrushenkov,~A.; Rauhut,~G.
  \latin{et~al.}  {MOLPRO}, version 2012.1, a package of ab initio programs.
  \emph{see http://www.molpro.net} \textbf{2012}, \relax
\mciteBstWouldAddEndPunctfalse
\mciteSetBstMidEndSepPunct{\mcitedefaultmidpunct}
{}{\mcitedefaultseppunct}\relax
\EndOfBibitem
\bibitem[Hollebeek \latin{et~al.}(1997)Hollebeek, Ho, and
  Rabitz]{Hollebeek1997}
Hollebeek,~T.; Ho,~T.-S.; Rabitz,~H. A fast algorithm for evaluating
  multidimensional potential energy surfaces. \emph{TJ. Chem. Phys.}
  \textbf{1997}, \emph{106}, 7223--7227\relax
\mciteBstWouldAddEndPuncttrue
\mciteSetBstMidEndSepPunct{\mcitedefaultmidpunct}
{\mcitedefaultendpunct}{\mcitedefaultseppunct}\relax
\EndOfBibitem
\bibitem[Unke and Meuwly(2017)Unke, and Meuwly]{Unke2017}
Unke,~O.~T.; Meuwly,~M. Toolkit for the construction of reproducing
  kernel-based representations of data: {A}pplication to multidimensional
  potential energy surfaces. \emph{J. Chem. Inf. Model.} \textbf{2017},
  \emph{57}, 1923--1931\relax
\mciteBstWouldAddEndPuncttrue
\mciteSetBstMidEndSepPunct{\mcitedefaultmidpunct}
{\mcitedefaultendpunct}{\mcitedefaultseppunct}\relax
\EndOfBibitem
\bibitem[Miller(1969)]{miller1969}
Miller,~W.~H. Coupled Equations and the Minimum Principle for Collisions of an
  Atom and a Diatomic Molecule, Including Rearrangements. \emph{The Journal of
  Chemical Physics} \textbf{1969}, \emph{50}, 407--418\relax
\mciteBstWouldAddEndPuncttrue
\mciteSetBstMidEndSepPunct{\mcitedefaultmidpunct}
{\mcitedefaultendpunct}{\mcitedefaultseppunct}\relax
\EndOfBibitem
\bibitem[Kendrick and Pack(1996)Kendrick, and Pack]{kendrick1996}
Kendrick,~B.; Pack,~R.~T. Geometric phase effects in {H+O}$_{2}$ scattering. I.
  Surface function solutions in the presence of a conical intersection.
  \emph{The Journal of Chemical Physics} \textbf{1996}, \emph{104},
  7475--7501\relax
\mciteBstWouldAddEndPuncttrue
\mciteSetBstMidEndSepPunct{\mcitedefaultmidpunct}
{\mcitedefaultendpunct}{\mcitedefaultseppunct}\relax
\EndOfBibitem
\bibitem[Voges \latin{et~al.}(2022)Voges, Gersema, Hartmann, Ospelkaus, and
  Zenesini]{voges2022}
Voges,~K.~K.; Gersema,~P.; Hartmann,~T.; Ospelkaus,~S.; Zenesini,~A. Hyperfine
  dependent atom-molecule loss analyzed by the analytic solution of few-body
  loss equations. \emph{Phys. Rev. Res.} \textbf{2022}, \emph{4}, 023184\relax
\mciteBstWouldAddEndPuncttrue
\mciteSetBstMidEndSepPunct{\mcitedefaultmidpunct}
{\mcitedefaultendpunct}{\mcitedefaultseppunct}\relax
\EndOfBibitem
\bibitem[Liu \latin{et~al.}(2025)Liu, Zhu, Luke, Babin, Gronowski, Ladjimi,
  Tomza, Bohn, Tscherbul, and Ni]{liu2025}
Liu,~Y.-X.; Zhu,~L.; Luke,~J.; Babin,~M.~C.; Gronowski,~M.; Ladjimi,~H.;
  Tomza,~M.; Bohn,~J.~L.; Tscherbul,~T.~V.; Ni,~K.-K. Hyperfine-to-rotational
  energy transfer in ultracold atom–molecule collisions of {Rb} and {KRb}.
  \emph{Nature Chemistry} \textbf{2025}, \emph{17}, 688--694\relax
\mciteBstWouldAddEndPuncttrue
\mciteSetBstMidEndSepPunct{\mcitedefaultmidpunct}
{\mcitedefaultendpunct}{\mcitedefaultseppunct}\relax
\EndOfBibitem
\bibitem[Takekoshi \latin{et~al.}(2014)Takekoshi, Reichs\"ollner, Schindewolf,
  Hutson, Sueur, Dulieu, Ferlaino, Grimm, and N\"agerl]{takekoshi2014}
Takekoshi,~T.; Reichs\"ollner,~L.; Schindewolf,~A.; Hutson,~J.~M.; Sueur,~C.
  R.~L.; Dulieu,~O.; Ferlaino,~F.; Grimm,~R.; N\"agerl,~H.-C. Ultracold dense
  samples of dipolar RbCs molecules in the rovibrational and hyperfine ground
  state. \emph{Phys. Rev. Lett.} \textbf{2014}, \emph{113}\relax
\mciteBstWouldAddEndPuncttrue
\mciteSetBstMidEndSepPunct{\mcitedefaultmidpunct}
{\mcitedefaultendpunct}{\mcitedefaultseppunct}\relax
\EndOfBibitem
\bibitem[Park \latin{et~al.}(2015)Park, Will, and Zwierlein]{park2015}
Park,~J.~W.; Will,~S.~A.; Zwierlein,~M.~W. Ultracold Dipolar Gas of Fermionic
  $^{23}\mathrm{Na}^{40}\mathrm{K}$ Molecules in Their Absolute Ground State.
  \emph{Phys. Rev. Lett.} \textbf{2015}, \emph{114}, 205302\relax
\mciteBstWouldAddEndPuncttrue
\mciteSetBstMidEndSepPunct{\mcitedefaultmidpunct}
{\mcitedefaultendpunct}{\mcitedefaultseppunct}\relax
\EndOfBibitem
\bibitem[Voges \latin{et~al.}(2020)Voges, Gersema, Meyer~zum Alten~Borgloh,
  Schulze, Hartmann, Zenesini, and Ospelkaus]{voges2020}
Voges,~K.~K.; Gersema,~P.; Meyer~zum Alten~Borgloh,~M.; Schulze,~T.~A.;
  Hartmann,~T.; Zenesini,~A.; Ospelkaus,~S. Ultracold Gas of Bosonic
  $^{23}\mathrm{Na}^{39}\mathrm{K}$ Ground-State Molecules. \emph{Phys. Rev.
  Lett.} \textbf{2020}, \emph{125}, 083401\relax
\mciteBstWouldAddEndPuncttrue
\mciteSetBstMidEndSepPunct{\mcitedefaultmidpunct}
{\mcitedefaultendpunct}{\mcitedefaultseppunct}\relax
\EndOfBibitem
\bibitem[Guo \latin{et~al.}(2018)Guo, Ye, He, Gonz\'alez-Mart\'{\i}nez, Vexiau,
  Qu\'em\'ener, and Wang]{guo2018}
Guo,~M.; Ye,~X.; He,~J.; Gonz\'alez-Mart\'{\i}nez,~M.~L.; Vexiau,~R.;
  Qu\'em\'ener,~G.; Wang,~D. Dipolar Collisions of Ultracold Ground-State
  Bosonic Molecules. \emph{Phys. Rev. X} \textbf{2018}, \emph{8}, 041044\relax
\mciteBstWouldAddEndPuncttrue
\mciteSetBstMidEndSepPunct{\mcitedefaultmidpunct}
{\mcitedefaultendpunct}{\mcitedefaultseppunct}\relax
\EndOfBibitem
\bibitem[Ye \latin{et~al.}(2018)Ye, Guo, González-Martínez, Quéméner, and
  Wang]{ye2018}
Ye,~X.; Guo,~M.; González-Martínez,~M.~L.; Quéméner,~G.; Wang,~D.
  Collisions of ultracold $^{23}${Na} $^{87}${Rb} molecules with controlled
  chemical reactivities. \emph{Science Advances} \textbf{2018}, \emph{4},
  eaaq0083\relax
\mciteBstWouldAddEndPuncttrue
\mciteSetBstMidEndSepPunct{\mcitedefaultmidpunct}
{\mcitedefaultendpunct}{\mcitedefaultseppunct}\relax
\EndOfBibitem
\bibitem[Gregory \latin{et~al.}(2019)Gregory, Frye, Blackmore, Bridge, Sawant,
  Hutson, and Cornish]{gregory2019}
Gregory,~P.~D.; Frye,~M.~D.; Blackmore,~J.~A.; Bridge,~E.~M.; Sawant,~R.;
  Hutson,~J.~M.; Cornish,~S.~L. Sticky collisions of ultracold {RbCs}
  molecules. \emph{Nature Communications} \textbf{2019}, \emph{10}, 3104\relax
\mciteBstWouldAddEndPuncttrue
\mciteSetBstMidEndSepPunct{\mcitedefaultmidpunct}
{\mcitedefaultendpunct}{\mcitedefaultseppunct}\relax
\EndOfBibitem
\bibitem[Gersema \latin{et~al.}(2021)Gersema, Voges, Meyer~zum Alten~Borgloh,
  Koch, Hartmann, Zenesini, Ospelkaus, Lin, He, and Wang]{gersema2021}
Gersema,~P.; Voges,~K.~K.; Meyer~zum Alten~Borgloh,~M.; Koch,~L.; Hartmann,~T.;
  Zenesini,~A.; Ospelkaus,~S.; Lin,~J.; He,~J.; Wang,~D. Probing Photoinduced
  Two-Body Loss of Ultracold Nonreactive Bosonic
  $^{23}\mathrm{Na}^{87}\mathrm{Rb}$ and $^{23}\mathrm{Na}^{39}\mathrm{K}$
  Molecules. \emph{Phys. Rev. Lett.} \textbf{2021}, \emph{127}, 163401\relax
\mciteBstWouldAddEndPuncttrue
\mciteSetBstMidEndSepPunct{\mcitedefaultmidpunct}
{\mcitedefaultendpunct}{\mcitedefaultseppunct}\relax
\EndOfBibitem
\bibitem[Bause \latin{et~al.}(2021)Bause, Schindewolf, Tao, Duda, Chen,
  Qu\'em\'ener, Karman, Christianen, Bloch, and Luo]{bause2021}
Bause,~R.; Schindewolf,~A.; Tao,~R.; Duda,~M.; Chen,~X.-Y.; Qu\'em\'ener,~G.;
  Karman,~T.; Christianen,~A.; Bloch,~I.; Luo,~X.-Y. Collisions of ultracold
  molecules in bright and dark optical dipole traps. \emph{Phys. Rev. Res.}
  \textbf{2021}, \emph{3}, 033013\relax
\mciteBstWouldAddEndPuncttrue
\mciteSetBstMidEndSepPunct{\mcitedefaultmidpunct}
{\mcitedefaultendpunct}{\mcitedefaultseppunct}\relax
\EndOfBibitem
\bibitem[Nichols \latin{et~al.}(2022)Nichols, Liu, Zhu, Hu, Liu, and
  Ni]{nichols2022}
Nichols,~M.~A.; Liu,~Y.-X.; Zhu,~L.; Hu,~M.-G.; Liu,~Y.; Ni,~K.-K. Detection of
  Long-Lived Complexes in Ultracold Atom-Molecule Collisions. \emph{Phys. Rev.
  X} \textbf{2022}, \emph{12}, 011049\relax
\mciteBstWouldAddEndPuncttrue
\mciteSetBstMidEndSepPunct{\mcitedefaultmidpunct}
{\mcitedefaultendpunct}{\mcitedefaultseppunct}\relax
\EndOfBibitem
\bibitem[Bause \latin{et~al.}(2023)Bause, Christianen, Schindewolf, Bloch, and
  Luo]{bause2023}
Bause,~R.; Christianen,~A.; Schindewolf,~A.; Bloch,~I.; Luo,~X.-Y. Ultracold
  Sticky Collisions: {T}heoretical and Experimental Status. \emph{The Journal
  of Physical Chemistry A} \textbf{2023}, \emph{127}, 729--741, PMID:
  36624934\relax
\mciteBstWouldAddEndPuncttrue
\mciteSetBstMidEndSepPunct{\mcitedefaultmidpunct}
{\mcitedefaultendpunct}{\mcitedefaultseppunct}\relax
\EndOfBibitem
\bibitem[Mayle \latin{et~al.}(2013)Mayle, Qu\'em\'ener, Ruzic, and
  Bohn]{Mayle2013}
Mayle,~M.; Qu\'em\'ener,~G.; Ruzic,~B.~P.; Bohn,~J.~L. Scattering of ultracold
  molecules in the highly resonant regime. \emph{Phys. Rev. A} \textbf{2013},
  \emph{87}, 012709\relax
\mciteBstWouldAddEndPuncttrue
\mciteSetBstMidEndSepPunct{\mcitedefaultmidpunct}
{\mcitedefaultendpunct}{\mcitedefaultseppunct}\relax
\EndOfBibitem
\end{mcitethebibliography}

\end{document}